\documentstyle[12pt]{article}

\input epsf.sty

\newcommand{\BEQ}{\begin{equation}}     
\newcommand{\BEA}{\begin{eqnarray}}
\newcommand{\EEQ}{\end{equation}}       
\newcommand{\EEA}{\end{eqnarray}}
\newcommand{\eps}{\varepsilon}          
\newcommand{\D}{{\rm d}}                
\newcommand{\II}{{\rm i}}               
\renewcommand{\Re}{{\rm Re\ }}          
\renewcommand{\Im}{{\rm Im\ }}          
\newcommand{\arcosh}{{\rm arcosh\,}}    
\newcommand{\zeile}[1]{\vskip #1 \baselineskip} 

\newcommand{\appsection}[2]{\setcounter{equation}{0} \section*{Annexe #1. #2}
\renewcommand{\theequation}{#1\arabic{equation}}
              \renewcommand{\thesection}{#1} }

\catcode`\@=11
\def\numberbysection{\@addtoreset{equation}{section}
        \def\theequation{\thesection.\arabic{equation}}}

\begin{document}

\begin{titlepage}

\vskip 1.5 cm
\begin{center}
{\Large \bf Sur la solution de Sundman du probl\`eme des 
trois corps}\footnote{{\rm Philosophia Scientiae} {\bf 5} (2), 
161--184 (2001);}
\end{center}

\vskip 2.0 cm
\centerline{  {\bf Malte Henkel}}
\vskip 0.5 cm
\centerline {Laboratoire de Physique des 
Mat\'eriaux,\footnote{Laboratoire associ\'e au CNRS UMR 7556} 
Universit\'e Henri Poincar\'e Nancy I,} 
\centerline{ B.P. 239, 
F -- 54506 Vand{\oe}uvre l\`es Nancy Cedex, France}
\begin{abstract}
\noindent 
Contrairement \`a une opinion largement r\'epandue, le probl\`eme
des trois corps poss\`ede une solution analytique. Cette solution
fut d\'ecouverte en 1909 par Sundman. Nous pr\'e\-sen\-tons dans cet
article les id\'ees de base et l'histoire de cette solution. \\ ~\\

\noindent 
Entgegen einer weitverbreiteten Meinung ist das Dreik\"orperproblem analytisch
l\"osbar. Diese L\"osung wurde 1909 von Sundman gefunden.
Die ihr zugrundeliegenden Ideen und ihre Geschichte 
werden in einfacher Form dargestellt.
\end{abstract}
\end{titlepage}

\section{Le probl\`eme des trois corps}

L'\'etude du  probl\`eme des trois corps a une longue 
histoire [Barrow-Green 1997, Diacu et Holmes 1996]. 
Comme beaucoup de probl\`emes qui ont suscit\'e un grand int\'er\^et, 
il se formule tr\`es facilement mais sa simplicit\'e
apparente cache une ph\'enom\'enologie tr\`es riche. 
Consid\'erons un ensemble de $n$ particules ponctuelles 
qui exercent des forces gravitationnelles l'un sur 
l'autre. Admettons qu'elles
ont les masses $m_i$, o\`u $i=1,2,\ldots,n$ et que leurs positions 
sont d\'ecrites \`a l'aide des vecteurs $\vec{r}_i(t)$ en fonction du temps
$t$. Les \'equations newtoniennes du mouvement s'\'ecrivent
\BEQ \label{Gl:dreiK}
m_i \frac{\D^2 \vec{r}_i(t)}{\D t^2} = - G \sum_{\stackrel{j=1}{j\ne i}}^{n} 
\frac{m_i m_j}{\left( \vec{r}_i(t) - \vec{r}_j(t)\right)^2}\cdot
\frac{\vec{r}_j(t) - \vec{r}_i(t)}{\left|\vec{r}_j(t) - \vec{r}_i(t)\right|} 
\;\; ; \;\; i=1,2,\ldots,n
\EEQ
o\`u $G$ est la constante gravitationnelle. Le probl\`eme des trois
corps s'obtient comme cas sp\'ecial $n=3$ 
du probl\`eme g\'en\'eral des $n$ corps
(\ref{Gl:dreiK}). 
En admettant des conditions initiales pour les
positions $\vec{r}_{i,0} := \vec{r}_i(0)$ et les vitesses 
$\vec{v}_{i,0} :=\D\vec{r}_i/\D t(0)$ on s'int\'eresse \`a calculer les
positions $\vec{r}_{i}(t)$, \`a partir des \'equations diff\'erentielles
(\ref{Gl:dreiK}), pour tous les temps $t$. Ainsi pos\'e, le probl\`eme est 
math\'ematiquement compl\`etement d\'efini et sa solution n'utilise que des
techniques math\'ematiques relevant d'un cours de m\'ecanique 
classique.\footnote{Des exemples physiques classiques 
du probl\`eme des $n$ corps sont \'evidemment fournis par le mouvement des 
plan\`etes autour du soleil, ou par le passage d'une com\`ete dans le syst\`eme
solaire. Des exemples plus r\'ecentes consid\`erent les orbites des
satellites artificiels, ou des plan\`etes r\'ecemment d\'ecouvertes autour
d'autres \'etoiles que le soleil.}

Quelques solutions particuli\`eres sont bien connues. Euler (1765) traite le
cas o\`u les trois corps sont align\'es. Lagrange (1772) resout le cas o\`u
les trois corps (dont un de masse n\'egligeable) 
sont aux sommets d'un triangle \'equilat\'eral rigide. 
Finalement, dans la solution de Moore (1993) et de
Chenciner et Montgomery (2001) trois corps de masse \'egale parcourent un huit.
Pour des d\'etails compl\'ementaires, on peut consulter [Montgomery 2001]. 

De part \`a sa difficult\'e technique consid\'erable, le probl\`eme des trois
corps suscite des commentaires souvent approximatifs et parfois erronn\'es. 
Par exemple, on lit souvent que ce
probl\`eme est insoluble et on attribue \`a Poincar\'e la d\'ecouverte
d'un tel th\'eor\`eme. Cette insolubilit\'e,
reli\'ee \`a la divergence des s\'eries que l'on utilisait pour r\'esoudre les
\'equations (\ref{Gl:dreiK}), aurait stimul\'e Poincar\'e pour d\'evelopper 
ses nouvelles m\'ethodes de nature topologique en m\'ecanique analytique
[Abraham-Marsden 1978, xvii]. 
Sous cette forme simple, ces on-dits ne sont que partiellement vrais
(et donc, en toute rigueur, enti\`erement faux). 
En effet, Sundman [1909] a donn\'e, dans un sens tr\`es pr\'ecis, une
solution exacte et g\'en\'erale du probl\`eme des trois corps.  
Elle s'\'ecrit \`a l'aide de s\'eries convergentes pour 
toute valeur (r\'eelle) du temps $t$. 

On peut se demander pourquoi un tel r\'esultat n'est connu qu'aux 
sp\'ecialistes. 
Cette note, destin\'ee \`a une audience non sp\'ecialis\'ee dans les
m\'ethodes modernes de la m\'ecanique analytique, 
pr\'esentera dans la section 2
un tr\`es bref r\'esum\'e historique dans lequel nous tenterons de 
rectifier les erreurs de certains commentaires et d'en expliquer l'origine. 
Dans la section 3 nous
pr\'esenterons un expos\'e des id\'ees essentielles de la solution de Sundman,
en suivant largement un article p\'edagogique de Saari [1990]. Des
conclusions dans la section 4 termineront cette note. En particulier, 
nous souligenerons le fait que la solution de Sundman, malgr\'e son 
int\'er\^et th\'eorique fondamental, n'est pas capable
de fournir des renseignements pratiques sur le comportement physique du 
syst\`eme. Quelques annexes d'intention p\'edagogique 
fournissent des compl\'e\-ments math\'e\-ma\-ti\-ques d'une 
nature un peu plus technique.

\section{Quelques remarques historiques} 

L'histoire du probl\`eme des trois corps, \`a la fin du 19e si\`ecle et 
au d\'ebut du 20e si\`ecle, est \'etroitement li\'ee \`a l'histoire
du journal {\it Acta Mathematica}, fond\'e et r\'edig\'e par G. Mittag-Leffler 
\`a Stockholm. Pour plus de d\'etails sur les relations entre Mittag-Leffler 
et Poincar\'e, on peut consulter [Nabonnand 1999a, 1999b]; 
pour une vision d'ensemble de l'histoire du probl\`eme des trois corps 
\`a la fin du 19e si\`ecle, on peut consulter [Barrow-Green 1997] et 
[Diacu et Holmes 1996].  

Mittag-Leffler a fait ses \'etudes en Allemagne, chez K. Weierstra{\ss}
\`a Berlin, chez E. Schering \`a G\"ottingen, 
et en France, chez C. Hermite \`a Paris. A cette \'epoque, 
caract\'eris\'ee par des nationalismes gratuits et des tensions politiques
franco-allemandes, les contacts et la connaissance r\'eciproque des
math\'ematiciens fran\c{c}ais et allemands \'etaient difficiles. Mittag-Leffler,
reconnu en particulier pour ses r\'esultats sur des fonctions m\'eromorphes qui 
portent aujourd'hui son nom [Knopp 1971], fut nomm\'e en 1881 professeur 
\`a Stockholm [Nabonnand 1999b, 71]. 
Vue la situation politique qui rendait d\'elicate la publication des travaux des
math\'ematiciens fran\c{c}ais dans les journaux allemands (et reciproquement),
il tente de tirer profit de ses bonnes relations avec les math\'ematiciens
fran\c{c}ais et allemands en cr\'eant un nouveau journal, nomm\'e
{\it Acta Mathematica}, qui publiera principalement des articles en fran\c{c}ais
et en allemand, dans le but avou\'e d'offrir 
 un forum international o\`u des travaux
d'importance peuvent \^etre publi\'es et lus dans les deux pays 
sans heurter les sensibilit\'es
politiques. En mars 1882, il demande \`a 
Poincar\'e de ``faire le succ\`es''\footnote{Lettre de Mittag-Leffler \`a 
Poincar\'e du 29 mars 1882} 
du journal et de lui envoyer pour publication ses m\'emoires
sur les fonctions fuchsiennes dont des 
\'enonc\'es de r\'esultats furent d\'ej\`a
publi\'es dans les {\it Comptes Rendus}. Poincar\'e accepte et son 
travail [Poincar\'e 1882]  
inaugure le premier volume des {\it Acta} en 1882. Il sera suivi d'une s\'erie 
de neuf autres {\oe}uvres dans les 10 premiers volumes des {\it Acta}.  
L'entreprise de Mittag-Leffler
est soutenue par ses coll\`egues scientifiques et le journal acquiert
rapidement une bonne r\'eputation internationale, [Nabonnand 1999b, 92].
Cependant, la situation financi\`ere du journal reste d\'elicate, et les
{\it Acta Mathematica} ne peuvent d\'emarrer 
que gr\^ace aux contributions de la 
fortune personnelle de Mittag-Leffler et des subventions du parlement su\'edois
(en 1895, la r\'eduction de la subvention par le parlement de
Stockholm jettera les {\it Acta Mathematica} dans une grave crise 
financi\`ere\footnote{Lettre de Mittag-Leffler \`a Poincar\'e du 20 avril 
1895.}). 

Pour toutes ces raisons, Mittag-Leffler cherche \`a augmenter le prestige
de son journal d\`es son lancement. 
Une telle opportunit\'e se pr\'esente \`a l'occasion
du 60e anniversaire du roi Oscar II de Su\'ede qui, ayant \'etudi\'e les
math\'ematiques lui-m\^eme, prend un int\'er\^et personnel 
dans les progr\`es des
math\'ematiques, et accepte de financer un prix pour la solution d'un
probl\`eme math\'ematique important. L'histoire et les \'ev\'enements
autour du prix du roi Oscar II ont \'et\'e d\'ecrits en d\'etail ailleurs
[Andersson 1994, Barrow-Green 1994, Barrow-Green 1997, Diacu 1996, 
Diacu et Holmes 1996, Nabonnand 1999a] et nous nous 
concentrons ici sur les aspects en lien direct avec la solution du 
probl\`eme par Sundman [1909, 1913] une trentaine d'ann\'ees plus tard.  

Le prix est annonc\'e, en fran\c{c}ais et en allemand, dans le volume 7
des {\it Acta} (et \'egalement dans de nombreux journaux internationaux
[Nabonnand 1999b]) et quatre questions sont propos\'ees. Voici la premi\`ere:
\begin{quotation}
{\it ``
1. \'Etant donn\'e un syst\`eme \ldots de points mat\'eriels qui s'attirent
mutuellement suivant la loi de Newton, on propose, sous la supposition qu'un 
choc de deux points n'ait jamais lieu, de repr\'esenter les coordonn\'ees de 
chaque point sous forme de s\'eries proc\'edant suivant quelques fonctions
connues du temps et qui convergent uniform\'ement pour toute valeur r\'eelle
de la variable. 

\ldots la solution \'etendra consid\'erablement nos connaissances par rapport
au syst\`eme du monde \ldots Lejeune-Dirichlet a communiqu\'e peu de temps 
avant sa mort \`a un g\'eom\`etre \ldots qu'il avait d\'ecouvert une m\'ethode
de l'int\'egration des \'equations diff\'erentielles \ldots {\rm [et]} 
il \'etait
parvenu \`a d\'emontrer d'une mani\`ere absolument rigoureuse la stabilit\'e
de notre syst\`eme plan\'etaire. Malheureusement nous ne connaissons rien
sur cette m\'ethode \ldots On peut pourtant supposer que cette m\'ethode
\'etait bas\'ee \ldots sur le d\'eveloppement d'une id\'ee fondamentale
et simple \ldots ''} [Mittag-Leffler 1885]
\end{quotation}

La suite est connue: 
Poincar\'e envoie\footnote{Tout laisse croire que les quatre 
questions avaient \'et\'e r\'edig\'ees de mani\`ere \`a pouvoir 
int\'eresser Poincar\'e. En effet, Mittag-Leffler
lui rappelle en juillet 1887 que le d\'elai final pour envoyer un manuscrit pour
le prix est le 1$^{\rm er}$ juin 1888 et ajoute: ``Si vous veuillez envoyer
quelque chose c'est guerre probable que quelqu'un vous d\'epassera.''}
le 17 mai 1888 son m\'emoire c\'el\`ebre [Poincar\'e 1890],\footnote{Cet 
article ne traite que du probl\`eme {\em restreint} des trois corps, 
c.\`a.d. les corps sont admis
\`a ne se d\'eplacer que dans un plan et $m_3 \ll m_1, m_2$.} 
inscrit sous la devise
{\em Nunquam praescriptos transibut sidera fines}, destin\'e au concours pour
le prix du roi Oscar II. Apr\`es d\'elib\'eration et de multiples voyages de
Mittag-Leffler, le comit\'e form\'e de Mittag-Leffler, Hermite et
Weierstra{\ss} [Nabonnand 1999b, 178-181]
d\'ecide de lui octroyer le prix, qui consiste en une
m\'edaille d'or et la somme de 2500 couronnes.\footnote{Pour comparaison, le
salaire annuel de Mittag-Leffler \`a Stockholm en 1881 \'etait de 7000 
couronnes. En 1894, 1 couronne \'etait \'equivalente \`a
1,40 francs fran\c{c}ais.}
Selon le r\`eglement, le m\'emoire gagnant doit \^etre publi\'e dans les
{\it Acta Mathematica}. En pr\'eparant cette publication, Poincar\'e 
fournit sur la demande de Mittag-Leffler des notes explicatives sur son 
m\'emoire (ce qui le rallonge d'une centaine de pages). En juillet 1889,
suite \`a une question de l'\'editeur associ\'e des {\it Acta}, E. Phragm\'en, 
Poincar\'e d\'ecouvre une erreur importante dans son 
m\'emoire.\footnote{Lettres de Poincar\'e \`a Mittag-Leffler du 16 juillet et 
du 1er d\'ecembre 1889.} Une grande
partie des r\'esultats que Poincar\'e a cr\^u \'etablir, entre autres 
sur la stabilit\'e du syst\`eme plan\'etaire, ne sont plus valables. 
Travaillant sur les corrections n\'ecessaires, il d\'ecouvre ce qu'on appelle
aujourd'hui les {\em points homoclines} et met en \'evidence le premier exemple
d'un syst\`eme au comportement chaotique. Au moment de la 
d\'ecouverte de l'erreur, le volume contenant le m\'emoire erron\'e est
imprim\'e et quelques exemplaires sont d\'ej\`a sortis de l'imprimerie. 
Afin d'\'eviter un scandale, Mittag-Leffler d\'ecide\footnote{Lettre de 
Mittag-Leffler \`a Poincar\'e du 5 d\'ecembre 1889.}
de rappeler le volume complet, de le d\'etruire et de l'imprimer \`a nouveau. 
Poincar\'e doit assurer le co\^ut
de cette seconde impression, d'un montant de 3585 couronnes et 63 {\o}re. 
Finalement, son m\'emoire du prix du roi Oscar II lui servira de base pour
son grand {\oe}uvre sur les nouvelles m\'ethodes de la m\'ecanique c\'eleste 
[Andersson 1994, Barrow-Green 1994, 1997, 
Diacu et Holmes 1996, Nabonnand 1999b].

Ajoutons ici quelques commentaires:

1. On lit souvent dans les articles de vulgarisation 
que le probl\`eme des trois corps n'est pas r\'esoluble. 
Formellement, un {\bf syst\`eme alg\'ebriquement int\'egrable}\footnote{La 
terminologie technique d\'efinie dans cet article sera mise en gras.} 
se caract\'erise par l'existence de certaines quantit\'es
$I_k = I_k( \vec{r}_1(t), \ldots, \vec{r}_n(t))$, appel\'ees 
{\bf int\'egrales premi\`eres}, qui restent constantes tout au long
de l'\'evolution du syst\`eme. Par cons\'equent, elles ne d\'ependent que
des valeurs initiales des positions $\vec{r}_{i,0}$ et des vitesses
$\vec{v}_{i,0}$, $i=1,\ldots,n$, c'est \`a dire
\BEQ
I_k = I_k \left( \vec{r}_{1,0},\ldots,\vec{r}_{n,0},
\vec{v}_{1,0},\ldots,\vec{v}_{n,0} \right)
\EEQ
Si l'on conna\^\i t autant d'int\'egrales premi\`eres $I_k$ que de variables
ind\'ependantes (on en a $6n$ pour le probl\`eme des $n$ corps) et si les
$I_k$ sont des fonctions suffisamment simples (c.\`a.d. alg\'ebriques)
des $\vec{r}_{i,0}$ et $\vec{v}_{i,0}$, le probl\`eme de r\'esoudre les
\'equations diff\'erentielles (\ref{Gl:dreiK}) se r\'eduit \`a trouver la 
solution d'un syst\`eme d'\'equations alg\'ebriques. Ceci est un probl\`eme
consid\'erablement plus simple que celui de r\'esoudre les \'equations 
(\ref{Gl:dreiK}) directement.  

On conna\^{\i}t $10$ int\'egrales premi\`eres ind\'ependantes 
pour le probl\`eme des $n$ corps. Elles peuvent s'\'ecrire comme~:\\
(i) les coordonn\'ees $\vec{R}$ du centre de masses du syst\`eme, \\
(ii) les composantes de la quantit\'e de mouvement totale $\vec{P}$, \\
(iii) les composantes du moment cin\'etique du syst\`eme $\vec{L}$ et \\
(iv) l'\'energie totale $E$. \\
Les \'equations (\ref{Gl:dreiK}) sont des \'equations 
diff\'erentielles du second ordre, 
on a donc $2\cdot 3n= 6n$ variables ind\'ependantes. 
On peut toujours \'eliminer
$12 = 10 +2$ variables du probl\`eme, gr\^ace aux 10 
int\'egrales premi\`eres et $2$ par deux sym\'etries d\'ecouvertes
par Jacobi en 1843.\footnote{Une variable s'\'elimine en consid\'erant une
des positions ou des vitesses comme une variable ind\'ependante 
(au lieu du temps $t$) et l'autre s'\'elimine \`a l'aide de l'``{\it 
\'elimination des n{\oe}uds}'' qu'on trouvera dans les livres sur la 
m\'ecanique c\'eleste.}
En somme, on reste avec un syst\`eme d'\'equations diff\'erentielles
\`a $6n-12 = 6(n-2)$ variables. Il est clair que le probl\`eme des $n=2$ corps
est int\'egrable mais il faudrait trouver d'autres 
int\'egrales premi\`eres pour 
rendre les cas $n>2$ int\'egrables. Or, ceci est impossible: Bruns  
a d\'emontr\'e en 1887 que toute 
int\'egrale premi\`ere qui est une fonction {\em alg\'ebrique} des
positions $\vec{r}_i(t)$ et des vitesses $\D\vec{r}_i(t)/\D t$ 
est une fonction des dix 
int\'egrales d\'ej\`a connues [Bruns 1887]. Poincar\'e [1890] a  
g\'en\'eralis\'e ce th\'eor\`eme aux fonctions {\em uniformes} (voir annexe A) 
des positions et des vitesses.\footnote{Painlev\'e a
g\'en\'eralis\'e ceci en 1897/1900 en d\'emontrant qu'il n'existe aucune 
int\'egrale $I_k$ alg\'ebriquement ind\'ependante qui soit une fonction 
alg\'ebrique/uniforme des vitesses. Aujourd'hui, des th\'eor\`emes comme celui 
de Ziglin [1983] fournissent des conditions n\'ecessaires pour 
l'int\'egrabilit\'e. On peut en d\'eduire que le probl\`eme des trois corps, 
au voisinage de la solution sp\'eciale de Lagrange (1772), ne poss\`ede 
pas de syst\`eme complet d'int\'egrales premi\`eres qui soient des fonctions
m\'eromorphes des positions et des vitesses [Tsygintsev 2000].} 
La non-existence d'autres int\'egrales premi\`eres \'etablit la 
non-int\'egrabilit\'e du probl\`eme des $n\geq 3$ corps. Pourtant, ceci 
n'implique pas qu'une solution exprimable \`a l'aide 
des s\'eries n'existait pas. Le 
th\'eor\`eme de Bruns et Poincar\'e montre seulement que certaines m\'ethodes 
alg\'ebriques sont insuffisantes pour r\'esoudre le probl\`eme. 

2. Il est bien connu qu'il existe des s\'eries perturbatives capables
de repr\'esenter les solutions du probl\`eme avec une tr\`es grande pr\'ecision.
Pour simplifier consid\'erons le cas $n=3$. Imaginons qu'un des trois corps
est le soleil et les deux autres des plan\`etes. 
En premi\`ere approximation, on n\'eglige les forces gravitationnelles entre
les deux plan\`etes, parce que leur masse est beaucoup plus petite que celle du
soleil.\footnote{Par exemple, la masse du soleil est environ mille fois celle
de la plan\`ete la plus grande, Jupiter.} 
Dans cette approximation, l'\'eq. (\ref{Gl:dreiK})
se d\'ecompose en deux probl\`emes \`a deux corps et s'int\'egre. Ensuite,
on rajoute les forces entre les plan\`etes en les traitant comme une
petite perturbation de la solution approximative obtenue auparavant. 
On trouve une meilleure approximation ``proche'' de la premi\`ere. 
En r\'ep\'etant cette proc\'edure, on obtient des expressions pour les
$\vec{r}_i(t)$ sous forme d'une s\'erie. Le cas des mouvements p\'eriodiques
pr\'esente des difficult\'es techniques qui furent r\'esolues \`a la fin du 19e
si\`ecle par l'utilisation des s\'eries de 
Lindstedt et Gyld\'en.\footnote{Un exemple simplifi\'e de la m\'ethode de 
Lindstedt sera expos\'e en annexe C.} 
Poincar\'e [1890] d\'emontre la divergence g\'en\'erique des s\'eries 
de Lindstedt, mais souligne aussi que
\begin{quotation}
{\it ``\ldots les consid\'erations qui pr\'ec\`edent 
n'enl\`event rien au m\'erite
pratique des d\'eveloppements de M. Lindstedt. Ils ne convergent pas; donc 
ils ne peuvent donner une approximation ind\'efinie; mail ils peuvent donner
assez rapidement une approximation tr\`es grande et tr\`es suffisante
pour les besoins de la pratique. 

Je serais d\'esol\'e d'avoir jet\'e quelque discr\'edit sur ces s\'eries 
\ldots parce que je regarde la m\'ethode de M. Lindstedt comme l'une des
meilleures qui soit connues.''} [Nabonnand 1999b, 193 note 4]
\end{quotation}

3. Le d\'eveloppement recherch\'e pour le prix du roi Oscar II n'est pas une
s\'erie du type perturbatif. 
On y cherche un d\'eveloppement convergent des coordonn\'ees $\vec{r}_i(t)$
sous la forme $\sum_{\nu=0}^{\infty} a_{\nu} f(t)^{\nu}$. Il semble
connu depuis longtemps que les solutions devraient \^etre de cette forme, comme
il appara\^{\i}t dans une lettre de Weierstra{\ss} \`a Mittag-Leffler de 1883
\begin{quotation}
{\it ``Angenommen nun, es gehe die Bewegung der Art vor sich, da{\ss} niemals
zwei Punkte zusammentreffen, so sind 
{\rm [$\vec{r}_i$, $i=1,\ldots,n$]} eindeutige
analytische Funktionen von $t$, nicht blo{\ss} f\"ur reelle Werte dieser
Gr\"o{\ss}e, sondern auch f\"ur alle komplexen, in denen die zweite
Koordinate (der Faktor von ${\rm i}$) dem absoluten Betrage nach unter
einer gewissen Grenze liegt. \ldots Man wird schwerlich {\em a priori} die
Bedingungen ermitteln k\"onnen, die erf\"ullt sein m\"ussen, damit niemals
zwei Punkte zusammentreffen k\"onnen, man wird vielmehr dieselben als erf\"ullt
vorraussetzen m\"ussen \ldots Poincar\'e hat \ldots {\rm [dieses]} Theorem
ebenfalls hergeleitet, wenigstens unter Vorraussetzung des Newtonschen
Gesetzes und daraus die Folgerung gezogen, es sei m\"oglich, die Koordinaten
aller Punkte in konvergierende Reihen der Form
$\sum_{\nu=0}^{\infty} a_{\nu} \varphi(t)^{\nu}$ zu entwickeln, wo 
$\varphi(t)$ eine bestimme Funktion von $t$ ist. Dies ist leicht einzusehen. 
\ldots Aber man erh\"alt auf diese Weise nicht Aufschlu{\ss} dar\"uber, ob
die gemachte Vorraussetzung erf\"ullt ist oder nicht und es ist auch die
Form, in der sich die Ausdr\"ucke der Koordinaten darstellen, nicht der Natur
der zu beschreibenden Bewegungen angemessen.''}
[Nabonnand 1999b, 119]\footnote{L'orthographe de l'allemand est mis \`a jour.}
\end{quotation}
Apr\`es avoir re\c{c}u de Mittag-Leffler une copie de cette lettre, 
Poincar\'e remarque qu'en effet la forme $\varphi(t) = \tanh(\alpha t /2)$ 
n'est utilisable que dans le cas o\`u il n'y a pas de collision, mais dans ses 
propres \'etudes il travaille avec la fonction 
$\varphi(t) = \tanh(\alpha s(t) /2)$, o\`u $s(t)$ est une fonction monotone
de $t$ encore \`a d\'eterminer. Selon Poincar\'e~:\footnote{Lettre de
Poincar\'e \`a Mittag-Leffler du 22 mai 1883.} 
\begin{quotation}
{\it ``Les solutions de ce probl\`eme {\rm [de d\'etermination de $s(t)$]} 
sont en nombre infini \ldots
Il est clair que dans chaque cas particulier, il faut choisir la plus 
{\em zweckm\"a{\ss}ig}.\footnote{utile, pratique; 
en allemand dans l'original.} 
Or je ne crois pas que dans le cas de la 
M\'ecanique C\'eleste celle que j'ai donn\'ee soit la plus 
{\em zweckm\"a{\ss}ig}, je crois qu'il y a mieux \`a trouver.''} 
\end{quotation}
Il appara\^{\i}t que la possibilit\'e des collisions entre les corps est
l'obstacle principal pour d'\'etablir une solution analytique du probl\`eme. 

4. En 1909, Sundman [1909] d\'emontre qu'une solution sous forme
de s\'eries convergentes pour tout temps $t$ existe pour le probl\`eme des
trois corps. Voici l'\'enonc\'e de son th\'eor\`eme: 
\begin{quotation}
{\it ``Si les constantes des aires dans 
le mouvement des trois corps par rapport \`a
leur centre commun de gravit\'e ne sont pas toutes nulles, on peut trouver une
variable $\tau$ telle que les coordonn\'ees des corps, leurs distances
mutuelles et le temps soient d\'eveloppables en s\'eries convergentes suivant 
les puissances de $\tau$, qui repr\'esentent le mouvement pour toutes les
valeurs r\'eelles du temps, et cela quels que soient les chocs qui se
produisent entre les corps.''} [Sundman 1909, 3]
\end{quotation}
Sur l'invitation de Mittag-Leffler, Sundman [1913] publie une exposition
d'ensemble de ses travaux dans les {\it Acta Mathematica}. 
Constatons qu'aucune trace de ce r\'esultat ne se trouve dans la correspondance 
entre Mittag-Leffler et Poincar\'e. 
Le travail de Sundman [1907,1909,1913] fournit la solution 
recherch\'ee par la question 1, pos\'ee par 
Weierstra{\ss} pour le prix du roi
Oscar II une trentaine d'ann\'ees auparavant, au moins pour 
le cas $n=3$.\footnote{Sundman a trouv\'e la solution pour le cas $n=3$. 
Presque un si\`ecle plus tard, et \`a l'aide de techniques  diff\'erentes, 
Wang [1991] a pu obtenir un r\'esultat analogue pour toute valeur de $n$.} 

Nous verrons dans la section prochaine dans quelle mesure les id\'ees
de Weierstra{\ss} et de Poincar\'e sont proches de celles 
de Sundman. 

\section{La th\'eorie de Sundman}

Nous pr\'esentons les id\'ees principales de la solution de Sundman du
probl\`eme des trois corps.\footnote{Cette section est inspir\'ee de l'article 
p\'edagogique de Saari [1990].} Nous allons nous servir du probl\`eme des
deux corps afin d'illustrer les id\'ees et nous mentionnons les 
g\'en\'eralisations n\'ecessaires pour le probl\`eme des trois 
corps.\footnote{Pour une pr\'esentation math\'ematiquement compl\`ete, dans une
notation moderne, voir [Abraham et Marsden 1978, Siegel et Moser 1971].}

Il est bien connu que dans le probl\`eme des deux corps, le mouvement
est dans le plan perpendiculaire au vecteur du 
{\bf moment cin\'etique}\footnote{Pour une particule \`a la
position $\vec{r}$, de vitesse $\vec{v}$ et de masse $m$, le vecteur du
moment cin\'etique s'obtient comme $\vec{L} = m \vec{v} \wedge \vec{r}$,
o\`u $\wedge$ est le produit vectoriel. Pour un syst\`eme de $n$
particules, chacune avec un moment cin\'etique $\vec{L}_i$, o\`u 
$i=1,2,\ldots,n$, le moment cin\'etique total du syst\`eme est 
$\vec{L}=\vec{L}_1+\vec{L}_2+\ldots+\vec{L}_n$.} $\vec{L}$. 
Comme $\vec{L}$ est conserv\'e, son orientation et donc aussi le plan du 
mouvement des deux corps sont fix\'es. Il est pratique d'utiliser des
variables complexes $z= x + \II y$ afin de d\'ecrire le mouvement dans le plan
(voir annexe A pour la terminologie des nombres complexes). 

\begin{figure}
\centerline{\epsfxsize=3.5in\epsfbox
{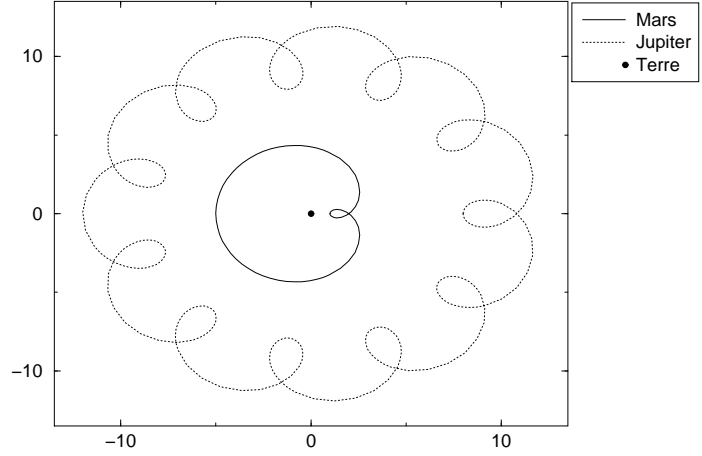}}
\caption{Mouvement de Mars (orbite int\'erieure) et de Jupiter 
(orbite ext\'erieure) vus depuis la terre (au centre).}
\label{fig2} \end{figure}

Afin d'en donner un exemple simple, consid\'erons le mouvement d'une plan\`ete
autour du soleil. Dans la figure~\ref{fig2},
nous montrons les orbites de Mars et de Jupiter, vues depuis la terre. Leur
mouvement apparent semble tr\`es compliqu\'e et il pourrait sember difficile 
de d\'ecrire et comprendre d'un point de vue physique une telle orbite. 
Bien entendu, nous savons aujourd'hui que cette difficult\'e n'est que 
apparente et qu'en r\'ealit\'e, les plan\`etes sont en orbite autour du 
soleil. Ces orbites sont en bonne approximation des cercles, dont les
rayons et les p\'eriodes sont rassembl\'es dans le tableau:

\begin{center}
\begin{tabular}{|l|cc|} \hline
plan\`ete & rayon & p\'eriode \\ \hline
Terre     & 2     & 1         \\
Mars      & 3     & 2         \\
Jupiter   & 10    & 12        \\ \hline
\end{tabular}
\end{center}

\noindent 
A l'aide des variables complexes, nous pouvons exprimer les orbites circulaires
des plan\`etes, vues depuis le soleil, sous la forme
\BEQ \label{gl:Helio}
z_{T}(t) = 2\, e^{2\pi\II t} \;\; ; \;\;
z_{M}(t) = 3\, e^{\pi\II t} \;\; ; \;\;
z_{J}(t) = 10\, e^{\pi\II t/6}
\EEQ
Le mouvement relatif \`a la 
terre s'obtient par un simple changement de rep\`ere.
Par exemple, pour l'orbite apparente de Mars, on a
\BEA
z(t) &=& z_{M}(t) - z_{T}(t) \nonumber \\
     &=& 3\, e^{\pi\II t} - 2\, e^{2\pi\II t} \nonumber \\
     &=& 2 + e^{\pi\II t} \left( 3 - 4 \cos(\pi t)\right)
\EEA
et g\'eom\'etriquement, l'orbite est un lima\c{c}on. De m\^eme, pour Jupiter, 
on obtient un lima\c{c}on relatif \`a un cercle tournant
\BEA
\tilde{z}(t) &=& z_{J}(t) - z_{T}(t) \nonumber \\
             &=& 2\, e^{-5\pi\II t/3} 
  + e^{\pi\II t/6}\left( 10 - 4 \cos\left(\frac{11\pi}{6}t\right)\right)
\EEA
Une telle description g\'eom\'etrique est d\'ej\`a utilis\'ee par
Ptolem\'ee et les astronomes de son \'ecole. Dans notre terminologie, il admet 
que les orbites apparentes sont de la forme
\BEQ \label{Gl:Ptol}
z_{\rm apparent}(t) = a_1\, e^{b_1\pi\II t} + a_2\, e^{b_2\pi\II t}
\EEQ
o\`u les $a_1,a_2,b_1,b_2$ sont des constantes \`a d\'eterminer. Le 
premier terme d\'ecrit le mouvement d'un plan\`ete selon le ``d\'ef\'erent'' et
le second d\'ecrit le mouvement additionnel selon ``l'\'epicycle''. Plus
syst\'ematiquement, le travail des astronomes avant Copernic (Kopernikus) 
revient \`a chercher une orbite plan\'etaire sous la forme
\BEQ
z_{\rm apparent}(t) = \sum_{j=1}^{\infty} a_{j}\, e^{b_{j}\pi\II t}
\EEQ
et de trouver les $a_{j}$ et $b_{j}$, c.\`a.d. d'effectuer une analyse de
Fourier d'une orbite plan\'etaire. Il est clair qu'en incluant suffisamment de
termes, on peut arrive \`a un tr\`es bon accord avec les observations. 
Bien que ce jugement soit compl\`etement r\'ecursif et anhistorique, on ne
peut s'emp\^echer de comparer la simplicit\'e des expressions lorsque l'on
choisit comme origine du rep\`ere le soleil (\ref{gl:Helio}) et la complication,
enti\`erement artificielle, d'expressions comme celles qui apparaissent 
dans (\ref{Gl:Ptol}). Un ``bon'' choix de variables permet de
simplifier consid\'erablement l'analyse des orbites apparemment
compliqu\'ees de la figure~\ref{fig2}.

De m\^eme, pour le probl\`eme des trois corps, c'est un nouveau 
choix de variables qui va 
permettre \`a Sundman de contourner un obstacle majeur \`a la compr\'ehension 
de ce probl\`eme. La th\'eorie de Sundman
utilise le fait (voir annexe A) que la valeur du rayon de convergence d'une
s\'erie est reli\'ee aux singularit\'es des fonctions complexes. 
Or, dans le cadre du probl\`eme des trois corps, l'interpr\'etation 
de telles singularit\'es est 
particuli\`erement simple~: En effet Painlev\'e a d\'emontr\'e en 1895 que 
pour $n=3$ toute singularit\'e des
fonctions $\vec{r}_i(t)$ par rapport \`a $t$ correspond soit (i) 
une collision entre deux corps soit (ii) une collision entre les trois 
corps.\footnote{Pour $n\geq 5$, des singularit\'es qui ne proviennent pas de
collisions simples peuvent appara\^{\i}tre. Pour $n=4$, on ne sait pas s'il 
existe d'autres singularit\'es que des collisions ou non.} 
Afin d'\'etablir l'existence d'une
solution sous forme d'une s\'erie convergente, 
Sundman [1907] part de ce th\'eor\`eme et progresse en deux \'etapes:
\begin{enumerate}
\item \'Eliminer les collisions {\em binaires} entre deux particules.
\item Exclure les collisions {\em ternaires} entre toutes les trois particules. 
\end{enumerate}
Dans le cas des {\em collisions binaires}, 
on peut se restreindre au probl\`eme \`a deux corps.\footnote{Lors d'une
collision binaire, le 3e corps est loin des deux autres qui se rencontrent. 
Au voisinage de la collision, son influence peut donc \^etre n\'eglig\'ee.} 
En effet, dans ce cas l'orbite a la forme d'une ellipse qui s'\'ecrit en  
coordonn\'ees polaires sous la forme
$r(\theta)=a/(1-e\cos\theta)\simeq a(1+e\cos\theta)$ o\`u $a$ est le 
{\bf demi-grand axe} et $e$ est l'{\bf excentricit\'e}
de l'orbite. Dans le cas $r=0$ et lorsque l'excentricit\'e $e$ tend vers 1, 
la variable angulaire $\theta$ passe d'une valeur proche de $0$ \`a une valeur 
proche de $2\pi$. Pour obtenir une solution analytique, cette singularit\'e 
en $\theta$ doit \^etre \'elimin\'ee.
Pour cela, on d\'ecrit le mouvement dans le plan \`a l'aide d'une variable
complexe $z=z(t)=x+\II y$. L'\'equation du mouvement s'\'ecrit alors sous
la forme
\BEQ
\frac{\D^2 z}{\D t^2} =-\frac{z}{r^3} \;\; , \;\; r = |z|
\EEQ 
On obtient une r\'egularisation du probl\`eme par 
le changement de variables [Sundman 1909, \'eq. (11)]
\BEQ \label{Gl:Sundman}
w = w(s) = z(t)^{1/2} \;\; , \;\; \D s = \frac{\D t}{r(t)}
\EEQ
Avec ces nouvelles variables, l'\'equation du mouvement s'\'ecrit
\BEQ \label{Gl:harmo}
\frac{\D^2 w(s)}{\D s^2} - \frac{h}{2} w(s) = 0
\EEQ
o\`u la constante $h$ correspond \`a l'\'energie 
totale.\footnote{Les calculs d\'etaill\'es sont
expos\'es dans l'annexe B. Nous montrerons que ce changement de variables 
est tr\`es naturel d'un point de vue physique.}  

La solution de l'\'equation (\ref{Gl:harmo}) est pour $h\ne 0$: 
$w(s) = w_{0} \cosh\left(\sqrt{h/2}\,s\right) + 
w_{1} \sinh\left(\sqrt{h/2}\,s\right)$ o\`u 
$w_{0,1}$ sont des constantes. Elle est sans singularit\'e et se d\'eveloppe
en s\'erie convergente en $s$ pour tout $|s|<\infty$.  

Pour le probl\`eme des trois corps l'id\'ee 
reste la m\^eme mais le temps $t$ est remplac\'e par une nouvelle variable 
$\omega$ [Sundman 1909, 1913]
\BEQ \label{Gl:dreiomega}
\D t = \left( 1 - e^{-r_0/\ell} \right) \left( 1 - e^{-r_1/\ell} \right)
 \left( 1 - e^{-r_2/\ell} \right) \D \omega
\EEQ
o\`u $r_{0,1,2}$ sont les distances entre les trois corps et $\ell$ est une 
constante. 

Le cas des {\em collisions ternaires}  semble \^etre 
fournie par le fait qu'une collision entre les trois particules n'est 
possible que si le moment cin\'etique $\vec{L}$ s'annule.\footnote{Cette 
proposition semble avoir 
d\'ej\`a \'et\'e \'etabli par Weierstra{\ss}, mais Sundman [1907] 
fut le premier \`a publier la d\'emonstration, 
voir aussi [Siegel et Moser 1971].} 
En effet, la condition $\vec{L}\ne \vec{0}$ 
est facile \`a contr\^oler \`a partir des conditions 
initiales\footnote{Si une collision ternaire se r\'ealise, on a 
$\vec{L}=\vec{0}$. Dans ce cas et au voisinage d'une collision, Sundman
trouve ``{\it que les corps se meuvent de telle mani\`ere 
\ldots {\rm [qu'ils]}  ou bien forment de plus en plus un triangle 
\'equilat\'eral ou bien se rangent de plus en plus en 
ligne droite.}'' [Sundman 1907] Ce type de configuration s'appelle 
{\em configuration centrale} et est aussi pr\'esent dans le
cadre du probl\`eme \`a $n$ corps.} et donc \'elimine toute 
singularit\'e en $\vec{r}_i(t)$ pour toute valeur r\'eelle de $t$.
Cependant, un examen plus pr\'ecis rev\`ele une complication technique 
dont nous allons discuter bri\`evement, avant de revenir \`a 
la solution de Sundman. 
 
Pour que le rayon de convergence du
d\'eveloppement en s\'erie du rayon vecteur $\vec{r}_i(t)$ soit non nul, 
il est n\'ecessaire qu'il n'y ait pas de singularit\'e au voisinage de l'axe 
r\'eel. Or ceci n'est pas trivial \`a \'etablir m\^eme dans le cas $n=2$. 
Dans ce cas, si l'excentricit\'e $e<1$, 
l'\'equation de l'orbite s'\'ecrit (\'equation de Kepler)
\BEQ
r(\upsilon) = a (1-e\cos\upsilon) \;\; , \;\;
t = \upsilon - e\sin \upsilon
\EEQ 
Bien que cette solution soit r\'eguli\`ere pour les temps r\'eels, 
la fonction $r(t)$ a des singularit\'es
complexes aux points 
\BEQ
\upsilon_k = 2\pi k + \II\, \arcosh (1/e)
\EEQ
(o\`u $k$ est un entier arbitraire) 
qui sont proches de l'axe r\'eel pour $e$ suffisamment proche de 1. 
Mais comme la valeur de $e$ d\'epend de celle
du moment cin\'etique $\vec{L}$ 
du probl\`eme des deux corps, on peut montrer lorsque $\vec{L}\ne \vec{0}$,
\BEQ
r> r_{\rm min}>0
\EEQ 
o\`u la valeur de $r_{\rm min}$ d\'epend de $\vec{L}$. 

\begin{figure}
\centerline{\epsfxsize=1.5in\ \epsfbox{
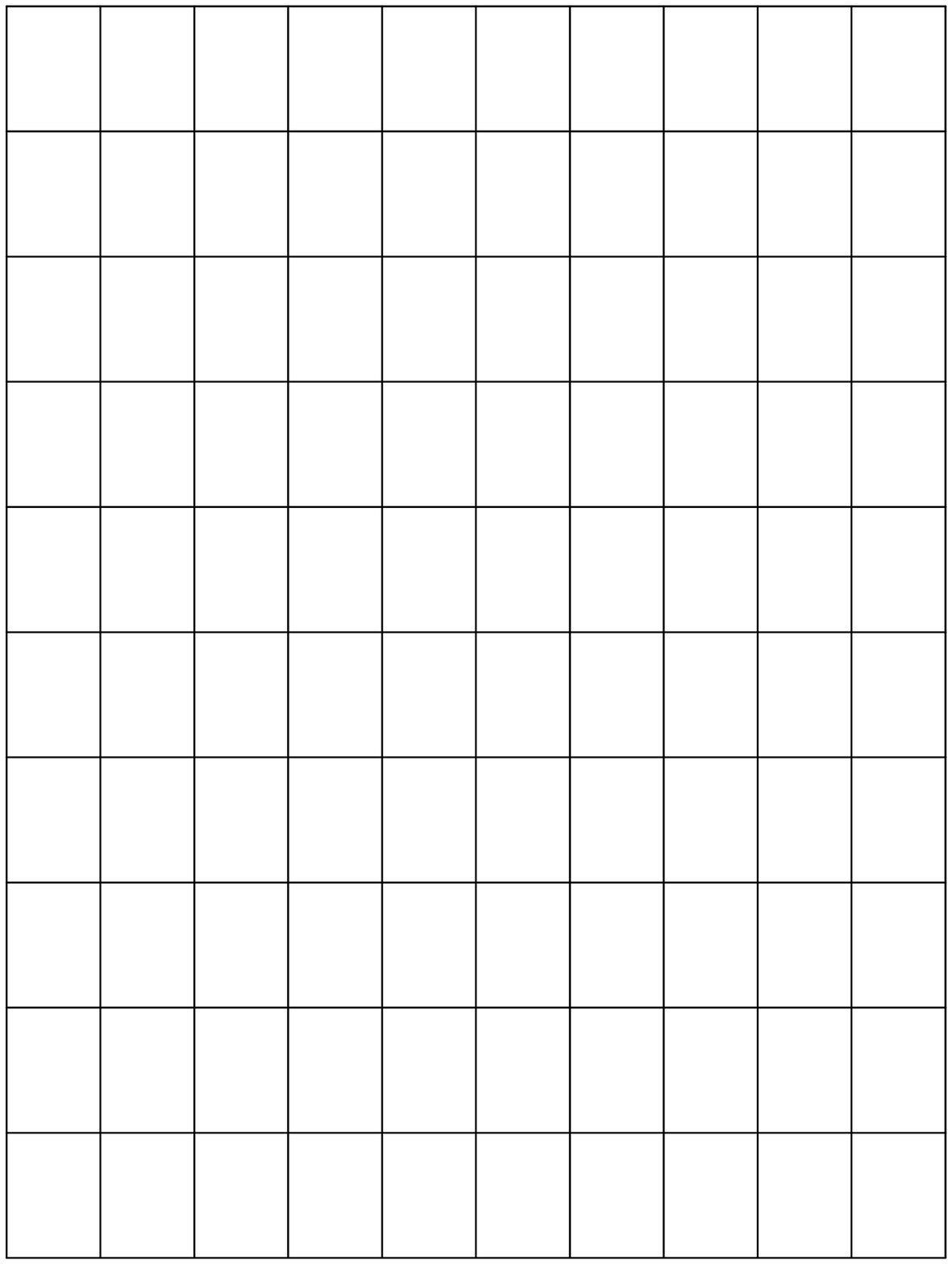}
\epsfxsize=1.5in\epsfbox{
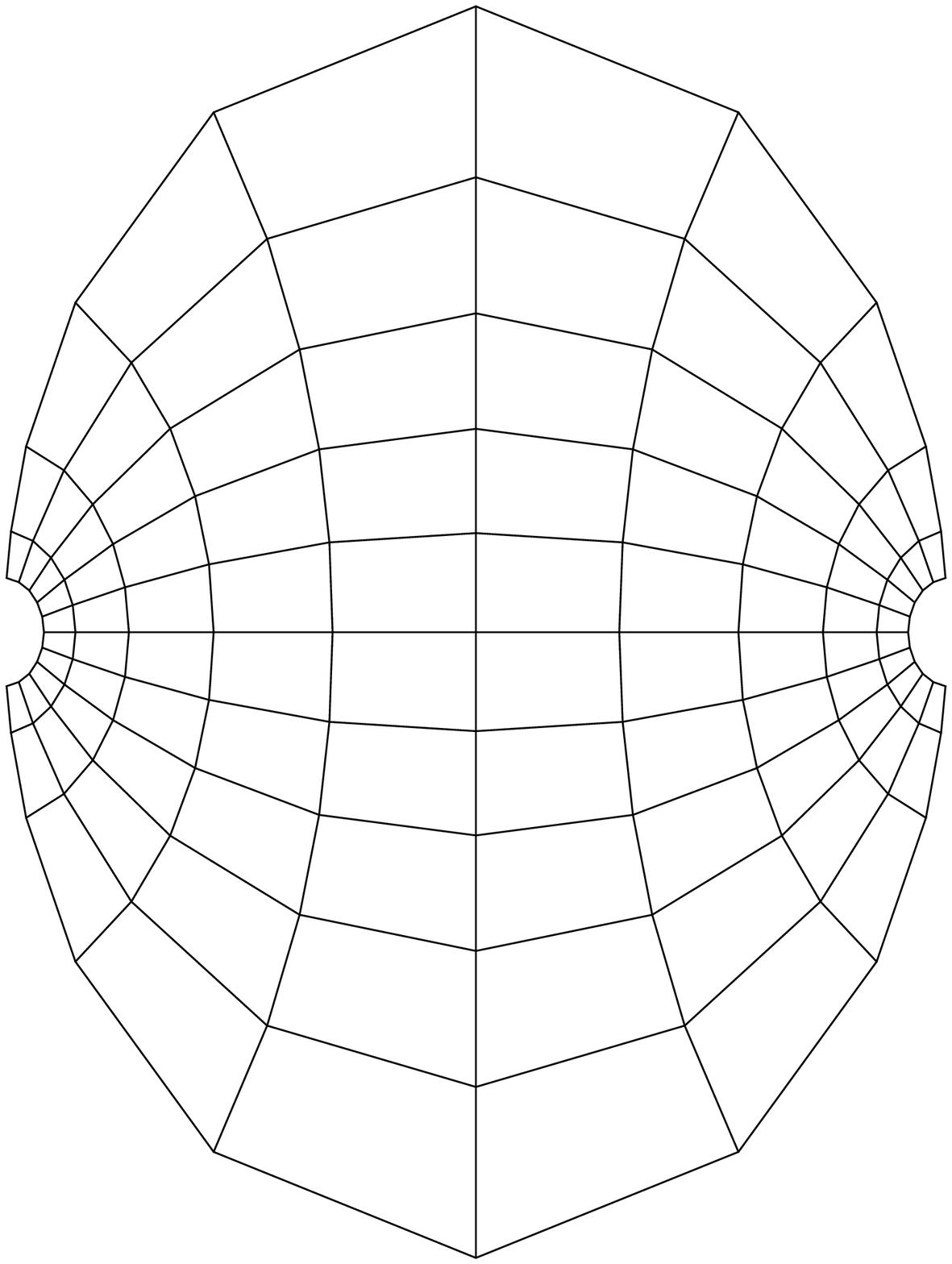}
}
\caption{Transformation conforme de la variable complexe $\omega$ (\`a
gauche) vers le plan complexe $\tau=\tanh\left(\pi\omega/4 \right)$ 
(\`a droite).}
\label{fig3} 
\end{figure}

Sundman [1907] a pu d\'emontrer un r\'esultat analogue pour
le probl\`eme des trois corps, c.\`a.d. que si $\vec{L}\ne\vec{0}$, on a
\BEQ
| \vec{r}_i(t) - \vec{r}_j(t)| > C(\vec{L}) > 0
\EEQ
o\`u $C(\vec{L})$ 
est une constante. Ayant ainsi \'etabli que pour 
$\vec{L}\ne \vec{0}$, il existe une borne inf\'erieure pour les distances entre
les corps, Sundman consid\`ere les temps formellement complexes et 
montre que le d\'eveloppement en s\'erie des rayons vecteurs
ne pr\'esente aucune singularit\'e dans un ruban $|\Im \omega| < B(L)$ 
o\`u $B(L)$ est une constante connue d\'ependant du moment cin\'etique 
$\vec{L}$ et $\omega$ est d\'efini par l'\'eq. (\ref{Gl:dreiomega}).
If suffit alors d'utiliser les transformations conformes  
consid\'er\'ees par Weierstra{\ss} et Poincar\'e 
\BEQ \label{gl:conf}
\tau = \tanh(\pi \omega/4B)
\EEQ
qui projette le ruban $|\Im \omega|< B$ sur l'int\'erieur du cercle $|\tau|<1$ 
(voir figure~\ref{fig3} pour le cas $B=1$). 
Dans la partie gauche, un morceau du plan complexe en $\omega$ est 
pr\'esent\'e. En exprimant les rayons vecteurs $\vec{r}_i = \vec{r}_i(\tau)$ en
fonctions de $t(\tau)$, il n'existe
aucune singularit\'e pour $|\tau|<1$. Comme $\tau=\pm 1$ correspond \`a 
$t=\pm\infty$, on a une repr\'esentation des positions sous forme d'une s\'erie
uniform\'ement convergente pour toute valeur $|t|<\infty$. 
Le probl\`eme des trois corps a donc trouv\'e sa solution sous la forme 
initialement conjectur\'ee par Weierstra{\ss}.  
 
\section{Remarques finales}

La solution de Sundman ne fait appel qu'aux techniques d'analyse standard 
des fonctions complexes, d\'ej\`a connues \`a son \'epoque. 
Dans ce sens, elle confirme l'attente de
Weierstra{\ss}, formul\'ee lors de l'annonce du prix du roi Oscar II, que la
solution devrait \^etre bas\'ee sur des id\'ees simples. On peut voir aussi
que par rapport \`a Weierstra{\ss} et Poincar\'e, Sundman ajoute deux nouvelles
id\'ees: (i) la r\'egularisation des collisions binaires \`a l'aide de son
changement de variables (\ref{Gl:dreiomega}) ce qui contourne la difficult\'e
de caract\'eriser le cas des collisions binaires et (ii) sa d\'emonstration de
s\'eparation minimale des trois corps lorsque $\vec{L}\ne\vec{0}$. 
Pourtant, la forme m\^eme de cette solution ne fournit aucune information 
sur le comportement du syst\`eme. 
Au contraire, le th\'eor\`eme de Sundman met en \'evidence que la propri\'et\'e
d'analyticit\'e  ne peut pas servir 
\`a caract\'eriser l'une ou l'autre propri\'et\'e qualitative 
d'un syst\`eme avec trois corps. En particulier,
aucun renseignement sur la stabilit\'e \`a des temps tr\`es longs n'est fourni. 
De plus, ces s\'eries sont inutiles pour des calculs num\'eriques. Certes, elles
convergent, mais le taux de convergence est tellement faible qu'un calcul
purement num\'erique, utilisant des techniques de r\'esolution directe
des \'equations de mouvement (\ref{Gl:dreiK}) sans faire appel aux 
d\'eveloppements des solutions en s\'eries, est plus 
fiable.\footnote{Ceci vient du fait que
le premier changement de variable (\ref{Gl:dreiomega}), de la forme
$\omega \sim t^{1/3}$, ralentit la dynamique. Enfin, le second changement
illustr\'e en figure \ref{fig3} concentre toute la dynamique au voisinage de 
$\tau\simeq \pm 1$.} Weierstra{\ss} et Poincar\'e, ont-ils pressenti ce
comportement en constatant que la variable $\tau$ \'etait 
``{\it nicht der Natur der zu beschreibenden Bewegung angemessen}''\footnote{
pas conforme \`a la nature du mouvement \`a d\'ecrire} et de ne 
pas \^etre ``{\it zweckm\"a{\ss}ig}'' ? 

Initialement, et surtout apr\`es la publication,  
de son article dans l'{\it Acta} [Sundman 1913], 
le travail de Sundman fut re\c{c}u avec grand 
int\'er\^et,\footnote{En 1913, Sundman recevait le prix Pont\'ecoulant 
de l'Academie des Sciences, dont la valeur a \'et\'e doubl\'ee \`a cette 
occasion [Barrow-Green 1997].} mais ce travail para\^{\i}t avoir \'et\'e 
tr\`es vite oubli\'e [Barrow-Green 1997]. 
Il n'est pas enti\`erement \'evident si cet oubli soit d\^u au 
succ\`es \'enorme des m\'ethodes
qualitatives de Poincar\'e pour l'\'etude des 
syt\`emes dynamiques ou que quelques unes
des id\'ees contenues dans les travaux de Sundman puissent encore aujourd'hui
stimuler de nouvelles directions de recherche.\footnote{Dans le
contexte de la m\'ecanique quantique, l'int\'egrabilit\'e du cas $n=2$ m\`ene 
\`a une sym\'etrie dynamique, que Pauli en 1927 a utilis\'e dans son 
traitement de l'atome d'hydrog\`ene. Existe-t-il une trace de la solution de
Sundman en m\'ecanique quantique, en tenant compte de la non-int\'egrabilit\'e
du cas $n=3$ ?} Siegel et Moser [1971]
consacrent tout un chapitre de leur trait\'e de la m\'ecanique celeste 
aux travaux de Sundman [1907,1909,1913] et soulignent leur
importance. En revanche, Abraham et Marsden [1978] ne mentionnent
qu'en passant les {\it ``{\oe}uvres classiques de Sundman (1913)''}.   

Notons encore que la solution de Sundman est constructive. En relation avec
le d\'ebat sur l'{\em intuitionisme} de Brouwer [Diacu 1996], 
il n'est pas sans int\'er\^et de remarquer que m\^eme une approche 
enti\`erement constructive ne donne pas toujours des r\'esultats directement 
utiles. Cet exemple\footnote{La solution satisfait \`a toutes les demandes
de rigueur math\'ematique, selon les standards \'etablies 
par Weierstra{\ss} et pourtant ce sont 
les m\'ethodes qualitatives de Poincar\'e et non elle qui nous renseignent 
sur le comportement, \'eventuellement chaotique, 
du probl\`eme des trois corps.} 
est une illustration excellente de la difficult\'e \`a 
d\'efinir des notions math\'ematiques d'une telle mani\`ere que des r\'esultats 
profonds peuvent \^etre obtenus. 
Dans le cadre de l'histoire du probl\`eme 
des trois corps, dont nous devons la compr\'ehension au progr\`es initi\'e 
par Poincar\'e [1890], et les d\'ebuts du journal
{\it Acta Mathematica}, cela fournit \'egalement une illustration
parfaite des occasions excellentes pour des d\'ecouvertes majeures 
qu'une approche scientifique r\'eellement internationale, 
d\'epassant entre autres des contraintes politiques, peut offrir.  

\zeile{2}\noindent {\large\bf Remerciements/Agradecimentos}\\

\noindent
Je remercie P. Nabonnand pour ses multiples commentaires et sa critique 
d\'etaill\'ee d'une premi\`ere version de cet article et 
T. Gourieux pour une lecture critique de ce travail.\\  
Agrade\c{c}o ao Complexo Interdisciplinar da Faculdade de 
Ci\^encias da Universidade de Lisboa pela sua hospitalidade, onde este 
trabalho foi escrito.

\appendix

\appsection{A}{Sur les fonctions complexes}

\begin{figure}
\centerline{\epsfxsize=5.0in\epsfbox
{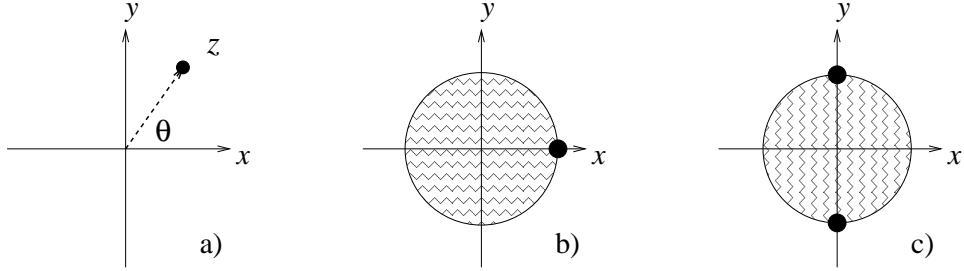}}
\caption{(a) La variable complexe $z=x+\II y$ et les cercles de convergence
et les singularit\'es des fonctions (b) $f_1(z)=1/(1-z)$ 
et (c) $f_2(z)=1/(1+z^2)$. Les deux s\'eries pour $f_1$ et $f_2$ autour de
$z_0=0$ convergent \`a l'int\'erieur du cercle $|z|=1$. }
\label{fig1} \end{figure}

Nous rappelons quelques faits \'el\'ementaires 
sur les fonctions analytiques $f(z)$ 
d'une variable complexe $z$ [Knopp 1976]. 
Un {\bf nombre complexe} $z= x+ \II y$ est
caract\'eris\'e \`a l'aide de deux nombres r\'eels $x, y$. On peut le 
repr\'esenter comme \'etant un point $(x,y)$ 
dans un plan, voir figure~\ref{fig1}a. 
$z$ redevient r\'eel si $y=0$. On appelle $x = \Re z$ la {\bf partie r\'eelle}
de $z$ et $y = \Im z$ la {\bf partie imaginaire} de $z$. 
L'{\bf unit\'e imaginaire} $\II$ satisfait \`a la condition $\II^2 = -1$. 
Le nombre $\bar{z} = x - \II y$ s'appelle le {\bf conjugu\'e complexe} de $z$. 
Alternativement, on peut \'ecrire en {\bf coordonn\'ees polaires} 
$z= \rho\, e^{\II \theta}$ ou $\theta$ est
l'angle indiqu\'e dans la figure~\ref{fig1} 
et $\rho = |z| = \sqrt{z\bar{z}}$ 
est la longueur de la ligne entre l'origine $(0,0)$ 
et le point $z$. Une {\bf fonction uniforme} $f$ est d\'efinie
en associant \`a chaque nombre complexe $z$ {\em un seul} 
autre nombre complexe $f(z)$. 
Exemples des fonctions uniformes 
sont des {\bf fonctions alg\'ebriques} comme $f(z)=1+z$ ou 
$(2+z)/(1+z^2)$ ou des {\bf fonctions enti\`eres} (c.\`a.d. analytiques dans le
plan complexe entier, sauf \`a l'infini $z=\infty$)
comme $e^{z}, \cos z$ ou $\sin z$. 
\`A cause de la relation d'Euler
\BEQ
e^{\II z} = \cos z + \II \sin z
\EEQ
les trois derni\`eres fonctions ne sont 
pas {\bf alg\'ebriquement ind\'ependantes}.
Il existent aussi des {\bf fonctions multiformes}, comme
$f(z) = \sqrt{z}$, o\`u \`a chaque valeur de $z$ correspondent plusieures
branches de $f(z)$ avec des valeurs diff\'erentes.

Comme on le fait pour des fonctions r\'eelles, on peut introduire la notion
de la {\bf d\'eriv\'ee} $f'(z) = \D f(z)/\D z$ d'une fonction complexe.
On dit que $f$ est {\bf analytique} dans un domaine $D$ du plan complexe 
si pour tout 
$z,z_0\in D$, $f'(z_0) =\lim_{z\to z_0} (f(z)-f(z_0))/(z-z_0)$ existe. 
Si $f$ est analytique dans un domaine autour d'un point $z_0$, il 
existe un nombre maximal $\rho>0$ et des constantes $f_n$ tels que 
$f(z)=\sum_{\nu=0}^{\infty} f_n (z-z_0)^n$ s'\'ecrit sous forme d'une 
s\'erie {\em uniform\'ement}\footnote{La d\'efinition pr\'ecise de la notion
de {\it convergence uniforme} d'une s\'erie comme utilis\'ee par Weierstra{\ss} 
et la diff\'erence avec la {\it convergence ponctuelle} se trouve dans tous les
textes d'analyse.} convergente pour $|z-z_0|<\rho$. $\rho$ s'appelle
{\bf rayon de convergence}. De plus, sur le cercle
$|z-z_0|=\rho$ il existe une singularit\'e o\`u $f$ n'est plus analytique. 
Un r\'esultat profond de la th\'eorie des fonctions complexes est qu'on peut 
caract\'eriser une fonction analytique par le lieu et la nature de ses
singularit\'es, voir [Knopp 1971]. 

Nous ne donnons ici que deux exemples illustratifs. Consid\'erons les
fonctions (qui fournissent des exemples des {\bf fonctions m\'eromorphes})
\BEQ
f_1(z) = \frac{1}{1-z} \;\; ; \;\;
f_2(z) = \frac{1}{1+z^2}
\EEQ
Si $|z|<1$, on peut repr\'esenter ces deux fonctions \`a l'aide des s\'eries
convergentes autour du point $z_0=0$ 
\BEQ
f_1(z) = \sum_{\nu=0}^{\infty} z^{\nu} \;\; ; \;\;
f_2(z) = \sum_{\nu=0}^{\infty} (-1)^{\nu} z^{2\nu}
\EEQ
A l'int\'erieur du cercle $|z|=1$ indiqu\'e dans la figure~\ref{fig1}, 
les s\'eries convergent uniform\'ement. Le domaine de convergence, 
qui est l'ensemble des $z$ complexes tels que $|z|<1$, est limit\'e par la
pr\'esence d'une o\`u plusieures singularit\'es sur le cercle $|z|=1$. 
Pour la fonction $f_1$, cette singularit\'e est visible comme une singularit\'e
sur l'axe r\'eel en $z=1$. En revanche, les singularit\'es de la fonction
$f_2$ sont en $z=\pm \II$, hors de l'axe r\'eel. M\^eme si $f_2(z)$ n'a 
aucune singularit\'e apparente pour toute valeur r\'eelle de $z$, ses 
singularit\'es complexes emp\^echent la s\'erie associ\'ee de converger pour
$|z|>1$. 

\appsection{B}{Sur les collisions binaires}

La notion de collision binaire dans le probl\`eme des deux corps est 
present\'ee de mani\`ere \'el\'ementaire. De part la conservation du 
vecteur du moment cin\'etique $\vec{L}$, le mouvement des deux corps est 
situ\'e est dans le plan perpendiculaire \`a $\vec{L}$. 
Les positions dans ce plan seront d\'efinies par les coordonn\'ees $(x, y)$ 
et la distance entre les deux corps par le module $r=|z|$ du nombre 
complexe $z = x+\II y$. Apr\`es un changement de coordonn\'ees,
l'\'equation du mouvement s'\'ecrit 
\BEQ \label{B:Bew}
\frac{\D^2 z}{\D t^2} =  - \frac{z}{r^3}
\EEQ  

La propri\'et\'e de {\em conservation de l'\'energie} s'\'etablit ainsi~:
pour $z$ et son conjugu\'e
complexe $\bar{z}$ on a $\D^2 z/\D t^2 =  - z / r^3$ et 
$\D^2 \bar{z}/\D t^2 =  - \bar{z} / r^3$. En multipliant la premi\`ere par
$\D\bar{z}/\D t$ et la seconde par $\D z/\D t$, la somme des deux \'equations
donne 
\BEQ
\frac{\D \bar{z}}{\D t}\frac{\D^2 z}{\D t^2} + 
\frac{\D z}{\D t}\frac{\D^2 \bar{z}}{\D t^2} =
-\frac{1}{r^3} \left( z \frac{\D \bar{z}}{\D t} + 
\bar{z} \frac{\D z}{\D t}\right)
\EEQ
Ceci s'\'ecrit aussi sous la forme
\BEQ
\frac{\D}{\D t} \left( \frac{\D z}{\D t}\frac{\D \bar{z}}{\D t}\right) 
= 2 \, \frac{\D}{\D t} \left( \frac{1}{\sqrt{ z \bar{z}\,}}\right)
\EEQ 
Par int\'egration, on trouve
\BEQ \label{Gl:Energie}
\left| \frac{\D z}{\D t} \right|^2 = 2\, \left( \frac{1}{r} + h \right)
\EEQ
o\`u la constante $h$ est l'\'energie totale du mouvement. 

L'\'equation (\ref{Gl:harmo}) se d\'eduit par changement de variables~: 
Dans la suite, nous \'ecrivons $\dot{z}=\D z/\D t$ et $z' = \D z/\D s$ et
de m\^eme pour $r$ et $w$. En appliquant le changement de variables 
(\ref{Gl:Sundman}) on obtient $\dot{z} = r^{-1} z'$ et $z'= 2 w w'$.
De plus, $r= |w|^2$ et la conservation de l'\'energie (\ref{Gl:Energie}) 
s'\'ecrit dans les nouvelles variables sous la forme
\BEQ \label{B:EnerW}
2 |w'|^2 = 1 + h r
\EEQ
\`A l'aide des relations $\ddot{z} = r^{-3} [ r w'' - r' z']$ et 
$\bar{w}^{-1} = w/r$, l'\'equation
du mouvement (\ref{B:Bew}) s'\'ecrit sous la forme 
\BEQ \label{B:BewW}
2w'' -\left(2|w'|^2-1\right)\frac{w}{r}=0
\EEQ
L'equation (\ref{Gl:harmo}) se d\'eduit imm\'ediatement de
(\ref{B:EnerW}) et (\ref{B:BewW}). 

Dans la suite de cette annexe, nous discutons la r\'egularisation du mouvement 
\`a travers une collision binaire. On peut toujours admettre\footnote{\`A l'aide
d'un changement de variables qui laisse invariant les \'equations de mouvement.}
que le mouvement est situ\'e sur l'axe x r\'eel. 
La distance entre les deux corps est d\'ecrite par 
l'\'equation du mouvement $\D^2 x/\D t^2 = -x^{-2}$. 
D'abord on s'int\'eresse \`a trouver
la fonction inverse $t=t(x)$ qui satisfait l'\'equation
\BEQ
x = (2+2hx) \left( \frac{\D t}{\D x}\right)^2
\EEQ
En int\'egrant directement, on obtient
\BEQ
t - t_0 = \int \!\D x\, \sqrt{\frac{x}{2+2hx}\,}
\EEQ
o\`u $t_0$ est une constante. 
L'int\'egrale se calcule, pour $h\ne 0$, \`a l'aide d'un changement 
de variables $x=h^{-1}\sinh^2 (u/2)$ et la solution finale s'\'ecrit 
(sous forme param\'etrique)
\BEQ \label{Gl:final}
\left( 2h\right)^{3/2} \left( t - t_0\right) = \sinh(u) - u \;\; , \;\;
2h \, x =  \cosh (u) -1 
\EEQ

Lors d'une collision, $x$ tend vers $0$. Au voisinage de 0, 
apr\`es l'\'equation (\ref{Gl:final}), on a  
$x \sim u^2$. De m\^eme, $\sinh(u)-u \sim u^3 \sim t-t_0$. 
On a au voisinage d'une collision
\BEQ
x \sim \left( t - t_0\right)^{2/3}
\EEQ
o\`u l'on identifie $t_0$ comme \'etant l'instant de la collision. 

La s\'erie $t-t_0 = u^3 \sum_{\nu=0}^{\infty} a_{\nu} u^{2\nu}$ est 
convergente pour tout $|u|<\infty$ et monotone en $u$ pour tout
$u$ r\'eel.\footnote{Ici, $a_{\nu} = (2h)^{-3/2}/((2\nu+3)!)$.} 
On en d\'eduit donc que la fonction
inverse $u=u(t-t_0)$ existe et se d\'eveloppe formellement en s\'erie 
$u = \left( t-t_0\right)^{1/3} 
\sum_{\nu=0}^{\infty} b_{\nu} \left( t-t_0 \right)^{2\nu/3}$
o\`u les $b_{\nu}$ s'obtiennent \`a partir des $a_{\nu}$. En injectant
ceci dans l'expression de $x(u)$ donn\'ee par l'\'equation (\ref{Gl:final}), 
on peut trouver des coefficients $B_{\nu}$ tels que
\BEQ \label{Gl:xReihe}
x = x \left( t-t_0\right) = \left( t-t_0\right)^{2/3}
\sum_{\nu=0}^{\infty} B_{\nu} h^{\nu}\left( t-t_0\right)^{2\nu/3}
\EEQ
Une telle s\'erie a un rayon de convergence non nul si la fonction complexe
qu'elle repr\'esente est d\'erivable. Consid\'erons alors la fonction complexe
$x=x(\tau)$ o\`u $\tau = (2h)^{-1}(\sinh(u) -u)^{2/3}$. 
Cette fonction est d\'efinie
si la fonction $\tau=\tau(u)$ a une fonction inverse $u=u(\tau)$. Pour ceci il
faut que la d\'eriv\'ee $\D \tau/\D u$ ne 'sannule pas. 
Or, $\D \tau/\D u = 0$ si
$u=u_k = 2\pi\II k$ o\`u $k =0, \pm 1, \pm 2,\ldots$ est un nombre entier 
arbitraire. Dans la suite, nous \'ecrivons 
$\tau_k = \tau(u_k) = (\sinh(u_{\pm k})-u_{\pm k})/(2h)$. Donc la d\'eriv\'ee
\BEQ
\frac{\D x}{\D\tau} = \frac{\D x}{\D u} \frac{\D u}{\D\tau} 
= \frac{\D x}{\D u} \left( \frac{\D\tau}{\D u}\right)^{-1} 
= \frac{3}{2} \cdot\frac{\sinh (u)\, [\sinh(u)-u]^{1/3}}{\cosh (u) -1}
\EEQ
existe si $u\ne u_k$. Autour de $u_0=0$, elle existe aussi, parce que
\BEQ
\lim_{u\to 0} \frac{\D x}{\D\tau} = \lim_{u\to 0} \frac{3}{2} u 
\left(\frac{u^3}{6}\right)^{1/3} 
\left(\frac{u^2}{2}\right)^{-1} = \frac{3}{\sqrt[3]{6}}
\EEQ
est une constante finie. Comme $u=0$ correspond 
\`a $\tau=0$ et donc \`a $t=t_0$, la fonction $x(\tau)$ est d\'erivable au
voisinage de la collision binaire et n'a aucune singularit\'e 
dans le domaine $|\tau|<|\tau_{\pm 1}|$. 
La s\'erie formelle (\ref{Gl:xReihe}) a donc un rayon de convergence 
$\rho \geq |\tau_{\pm 1}| >0$ et repr\'esente une fonction analytique. 

Si $u=u_k$, on a $x=0$ \`a l'instant $t$ donn\'e par $\tau=\tau_k$. Par
cons\'equent, les $u_k$ caract\'erisent une suite de (pseudo)collisions. 
Le passage de $t$ \`a $\tau$ nous a permis de r\'egulariser les \'equations
afin de pouvoir continuer analytiquement le mouvement au-del\'a la collision
en $t=t_0$. La non-analyticit\'e apparente \`a $\tau=\tau_k$ pour $k\ne 0$ 
indique simplement que le choix de $\tau$ n'est pas encore optimal et 
devrait \^etre remplac\'e par la variable $s$ d\'efinie par 
Sundman.\footnote{Pour le rapport avec les variables (\ref{Gl:Sundman}) 
utilis\'ees par Sundman, on a au voisinage d'une
collision $x(t)\sim (t-t_{0})^{2/3}$ et $u\sim (t-t_{0})^{1/3}$, donc
$\D u= \frac{1}{3} (t-t_{0})^{-2/3} \D t$. On retrouve ainsi une propri\'et\'e
de la r\'egularisation de Sundman dans (\ref{Gl:Sundman}).} 
 
Notons qu'on a aussi une analyticit\'e de $x$ en fonction de $h$. 
 
\appsection{C}{Un exemple de s\'erie de Lindstedt}

Nous exposons bri\`evement la m\'ethode perturbative de Lindstedt, en suivant
[Rand et Armbruster 1987]. A titre d'exemple, 
et dans un souci de simplicit\'e technique, 
nous consid\'erons {\bf l'oscillateur de van der Pol}, d\'efini par
l'\'equation diff\'erentielle non lin\'eaire
\BEQ \label{Gl:vanderPol}
\frac{\D^2 x}{\D t^2} + x + \eps (x^2 -1 ) \frac{\D x}{\D t} = 0
\EEQ
o\`u $x=x(t)$ est la fonction \`a d\'eterminer en fonction du param\`etre
$\eps$.\footnote{Si $\eps=0$, on retrouve l'\'equation d'un 
oscillateur harmonique.} Pour simplifier l'expos\'e, nous poserons comme
{\bf conditions initiales} $x(0)=\Xi_0$ et $\D x/\D t(0)=0$, o\`u $\Xi_0$
est une constante. On sait que pour toute valeur de $\eps$, 
il existe une solution p\'eriodique, appel\'ee {\bf cycle limite} 
et que $x(t)$ va toujours \'evoluer vers le cycle limite lorsque 
$t$ tend vers l'infini. Comme
seul le cas $\eps=0$ est facilement soluble, on se propose de r\'esoudre le cas
g\'en\'eral sous forme d'une {\bf s\'erie perturbative}
\BEQ
x(t) = x_0(t) + \eps\, x_1(t) + \eps^2\, x_2(t) + \cdots 
\EEQ
et d'utiliser un changement de variable temporelle $\tau = \omega t$
propos\'e par Lindstedt o\`u
\BEQ
\omega = 1 + k_1\, \eps + k_2\, \eps^2 + \cdots
\EEQ
En \'ecrivant $f'(\tau)$ pour la d\'eriv\'ee par rapport \`a $\tau$ de la
fonction $f(\tau)$, on obtient le syst\`eme d'\'equations, jusqu'aux termes
du second ordre en $\eps$ inclus
\BEA
x_0{''} + x_0 &=& 0 \nonumber \\
x_1{''} + x_1 &=& x_0{'} (1-x_0^2) -2k_1 x_0 \label{Gl:Rekurs} \\
x_2{''} + x_2 &=& x_1{'} (1-x_0^2) -2 x_0 x_0{'} x_1 -2 k_1 x_1{''}
-(2k_2 +k_1^2)x_0{''} + k_1 (1-x_0^2) x_0{'} \nonumber
\EEA
qui peut \^etre r\'esolu par r\'ecurrence. Les conditions initiales 
$\Xi_0 = X_0 + \eps X_1 + \eps^2 X_2 + \cdots$ se d\'eveloppent de la m\^eme
mani\`ere et on tient comme conditions initiales \`a l'instant $t=0$
\BEQ \label{Gl:Init}
x_i(0) = X_i \;\; , \;\; {x_i}'(0) =0 \;\; ; \;\; i = 0,1,2,\ldots
\EEQ
On d\'eduit des \'equations (\ref{Gl:Rekurs}), (\ref{Gl:Init}) que
$x_0(\tau) = X_0 \cos \tau$ et
\BEQ \label{Gl:C5}
x_1{''} + x_1 = \left[ \frac{1}{4} X_0^3 - X_0 \right] \sin \tau 
+ 2 k_1 X_0 \cos \tau + \frac{1}{4} X_0^3 \sin (3\tau)
\EEQ
La r\'esolution de (\ref{Gl:C5}) peut conduire \`a l'apparition
de termes dits s\'eculaires de la forme $\tau\sin\tau$ ou $\tau\cos\tau$. 
De tels termes croissent ind\'efiniment avec le
temps et rendent les d\'eveloppements non-convergents. Pour les \'eliminer, il
faut que les coefficients de $\sin\tau$ et $\cos\tau$ 
dans l'\'equation (\ref{Gl:C5}) s'annulent, d'o\`u $X_0=2$ et $k_1=0$. 
Par r\'ecurrence, on obtient 
\BEQ 
x_1(\tau) = -\frac{1}{4}\sin(3\tau) 
+\frac{3}{4}\cos(3\tau) + X_1\cos\tau
\EEQ
\begin{displaymath}
x_{2}{''} + x_2 = \left[4k_2+\frac{1}{4}\right]\cos\tau +2X_1 \sin\tau 
-\frac{3}{2}\cos(3\tau) +3X_1\sin(3\tau) +\frac{5}{4}\cos(5\tau)
\end{displaymath}
L'\'elimination des termes s\'eculaires am\`ene \`a poser\footnote{Le fait 
que $k_2\ne 0$ d\'emontre que $\omega\ne 1$ est n\'ecessaire.} 
$X_1=0$, $k_2=-\frac{1}{16}$.
Notons encore que l'amplitude et la fr\'equence $\omega$ de la solution
p\'eriodique sont compl\`etement fix\'ees par la non lin\'earit\'e
de l'\'equation (\ref{Gl:vanderPol}) et ne d\'ependent pas des 
conditions initiales. \`A partir des expressions 
explicites on voit aussi qu'on obtient le cycle limite
sous forme d'une {\bf s\'erie trigonom\'etrique} dont nous avons trouv\'e
les premiers termes
\BEQ
x(t) = 2 \cos(\omega t) + 
\frac{\eps}{4} \left[ 3 \cos(3\omega t)-\sin(3\omega t) \right] 
+ \cdots \;\; , \;\;
\omega = 1 - \frac{\eps^2}{16} + \cdots
\EEQ


{\small
\section*{Ref\'erences}
\begin{description}
\item Abraham, R. et Marsden, J.E. \\
1978~~ {\it Foundations of mechanics}, 2e \'edition, Benjamin (Reading, Mass.)
\item Andersson, K.G. \\
1994~~ {\it Poincar\'e's discovery of homoclinic points},
Archive Hist. Exact Sciences {\bf 48}, 133--147
\item Barrow-Green, J. \\
1994~~ {\it Oscar II's prize competition and the error in Poincar\'e's memoir
on the three-body problem},
Archive Hist. Exact Sciences {\bf 48}, 107--131 \\
1997~~ {\it Poincar\'e and the three-body problem}, 
American and London Mathematical Societies (London)
\item Bruns, E.H. \\
1887~~ {\it \"Uber die Integrale des Vielk\"orperproblems},
Acta Mathematica {\bf 11}, 25--96
\item Diacu, F.N. \\
1996~~ {\it The solution of the $n$-body problem},
Math. Intelligencer {\bf 18}, No. 3, 66--70
\item Diacu, F.N. et P. Holmes, P. \\
1996~~ {\it Celestial encounters -- the origins of chaos and stability}, 
Princeton University Press (Princeton)
\item Knopp, K. \\
1976~~ {\it Funktionentheorie I}, 13. Auflage, Walter de Gruyter (Berlin)\\ 
1971~~ {\it Funktionentheorie II}, 12. Auflage, Walter de Gruyter (Berlin) 
\item Mittag-Leffler, G. \\
1885~~ {\it Communication sur un prix de math\'ematiques fond\'e
par le roi Oscar II}, Acta Mathematica {\bf 7}, I--VI (1885/86)
\item Montgomery, R. \\
2001~~ {\it A New Solution to the Three-Body Problem},
Notices of the American Mathematical Society, {\bf 48}, 471 -- 481 
\item Nabonnand, P. \\
1999a~ {\it The Poincar\'e--Mittag-Leffler
relationship}, Math. Intelligencer {\bf 21}, No. 2, 58--64\\
1999b~ {\it La correspondance entre Henri
Poincar\'e et G\"osta Mittag-Leffler}, Birkh\"auser (Basel)
\item Poincar\'e, H. \\
1882~~ {\it Sur les groupes fuchsiens}, 
Acta Mathematica, {\bf 1}, 1 -- 62 (1882); {\OE}uvres 2, 108-168\\
1890~~ {\it Sur le probl\`eme des trois corps et les \'equations de la 
dynamique}, Acta Mathematica {\bf 13}, 1 -- 270 (1890)
\item Rand, R.H.  et Armbruster, D.\\
1987~~ {\it Perturbation Methods, Bifurcation Theory and Computer Algebra},
Springer (Heidelberg), ch. 1
\item Saari, D.G. \\
1990~~ {\it A visit to the newtonian $N$-body
problem via elementary complex variables}, Am. Math. Monthly {\bf 97},
105-119
\item Siegel, C.L. et Moser, J.K. \\
1971~~ {\it Lectures on Celestial Mechanics 
(Vorlesungen \"uber Himmelsmechanik)}, Springer (Heidelberg 1956/71). 
\item Sundman, K.F. \\
1907~~ {\it Recherches sur le probl\`eme des trois corps},
Acta Societatis Scientiarium Fennicae {\bf 34}, No 6\\
1909~~ {\it Nouvelles recherches sur le probl\`eme des trois corps},
Acta Societatis Scientiarium Fennicae, {\bf 35}, No 9\\
1913~~ {\it M\'emoire sur le probl\`eme des trois corps},
Acta Mathematica {\bf 36}, 105--179
\item Tsygvintsev, A. \\
2000~~ {\it La non-int\'egrabilit\'e m\'eromorphe du probl\`eme plan des 
trois corps}, C.R. Acad. Sci. Paris (S\'erie I) {\bf 331}, 241-244
\item Wang, Q.D. \\
1991~~ {\it The global solution of the $n$-body problem},
Celestial Mechanics and Dynamical Astronomy {\bf 50}, 73--88
\item Ziglin, S.L. \\ 
1983~~ {\it Branching of solutions and non-existence of first integrals in 
Hamiltonian Mechanics}, Funct. Anal. Appl. {\bf 16}, 181--189 et 
{\bf 17}, 6--17 
\end{description}
}

\end{document}